\documentclass[aps,pra,reprint,groupedaddress,showpacs]{revtex4-1}
\usepackage{graphicx,graphics}
\usepackage[table]{xcolor}
\usepackage{epstopdf}
\usepackage{dcolumn}
\usepackage{amsmath,amssymb,amsfonts}
\usepackage{latexsym,verbatim}
\usepackage{bm}
\usepackage{bbold}
\usepackage{color}
\usepackage[breaklinks=false,colorlinks,citecolor=blue,linkcolor=blue,urlcolor=blue]{hyperref}
\usepackage[abs]{overpic}
\usepackage{textcomp} 
\usepackage{mathtools}
\usepackage{braket}
\usepackage{comment}
\usepackage{tcolorbox}
\usepackage{soul}    
\usepackage[normalem]{ulem}
\usepackage[utf8]{inputenc}    
\usepackage[T2A]{fontenc}      
\usepackage[russian,english]{babel}

\definecolor{dkgreen}{rgb}{0.0, 0.5, 0.0}

\newif\ifdraft
\drafttrue 
\draftfalse 

\ifdraft

  \newcommand\FCM[1]{\textcolor{red}{#1}}
\else

  \newcommand\FCM[1]{}
\fi

\begin{document}

\title{Polarization dependence of spin-electric transitions in molecular exchange qubits}

\author{Filippo Troiani}
\email{filippo.troiani@nano.cnr.it}
\affiliation{Centro S3, CNR-Istituto di Nanoscienze, I-41125 Modena, Italy}
\author{Athanassios K. Boudalis}
\affiliation{Institut de Chimie (UMR7177 CNRS-Université de Strasbourg), CS-90032, F-67081 Strasbourg Cedex, France}

\begin{abstract} 
Quasi-optical experiments are emerging as a powerful technique to probe magnetic transitions in molecular spin systems. However, the simultaneous presence of the electric- and magnetic-dipole induced transitions poses the challenge of discriminating between these two contributions. Besides, the identification of the spin-electric transitions can hardly rely on the peak intensity, because of the current uncertainties on the value of the spin-electric coupling in most molecular compounds. Here, we compute the polarizations required for electric- and magnetic-dipole induced transitions through spin-Hamiltonian models of molecular spin triangles. We show that the polarization allows a clear discrimination between the two kinds of transitions. In addition, it allows one to identify the physical origin of the zero-field splitting in the ground multiplet, a debated issue with significant implications on the coherence properties of the spin qubit implemented in molecular spin triangles.
\end{abstract}

\date{\today}

\maketitle

\section{Introduction}

Spin-electric, or magnetoelectric (ME), phenomena have been considered for several decades now. Following the first discussion on the possibility of piezomagnetic and ME effects, \cite{Landau1958Electrodynamics} Dzyaloshinskii predicted their occurrence in $\text{Cr}_2\text{O}_3$, relying on symmetry considerations \cite{Dzyaloshinskii1960On}. The experimental confirmation was produced immediately hereafter \cite{Astrov1960magnetoelectric}, and successively corroborated by similar experiments \cite{Rado1962Magnetoelectric}. Roughly at that time, Electric Paramagnetic Resonance (EPR) experiments under static electric fields emerged as an alternative means for investigating spin-electric effects in diverse magnetic systems \cite{Mims1976linear}. Today, a renaissance of such studies is taking place, revolving around the possibility of electric control of molecular spins in the context of quantum applications \cite{Troiani11a,Moreno21a,Chiesa24a}. 

Electric fields can be used in different ways for manipulating molecular spins. One possibility is represented by the application of static fields, whose intensity is large enough to modify the eigenstates or the eigenvalues of the system, so as to modify its magnetic response \cite{Baadji09a,Boudalis18a,Lewkowitz23a}. Alternatively, baseband voltage pulses can be used for modulating the phase accumulation within spin-echo sequences in pulsed EPR experiments \cite{George2013Coherent,Liu19a,Liu2021Quantum,Robert19a}. In a different approach, moderate low-frequency electric fields induce a modulation of the magnetic couplings and of the resulting magnetic response of the system \cite{Fittipaldi19a,Kintzel21a,Cini25a}. On the other hand, weak and high-frequency electric fields can resonantly induce transitions between unperturbed system eigenstates. In polynuclear systems, this allows one to access spin degrees of freedom that are not amenable to magnetic manipulation \cite{leMardele25a}.
However, unlike the energies of the transitions between Zeeman levels, which are generally addressed in EPR experiments, those of the electric-field induced transitions are hardly tunable through the static magnetic field. This makes it more difficult to address them through cavities, and calls for the use of quasi-optical free-space techniques, such as magneto far-infrared spectroscopy \cite{Kragskow22a,Moseley20a,Blockmon21a}. Such an experimental approach also offers the possibility of addressing the system with radiation of different polarizations, an opportunity that we extensively explore hereafter.

\begin{figure}
\centering
\includegraphics[width=0.4\textwidth]{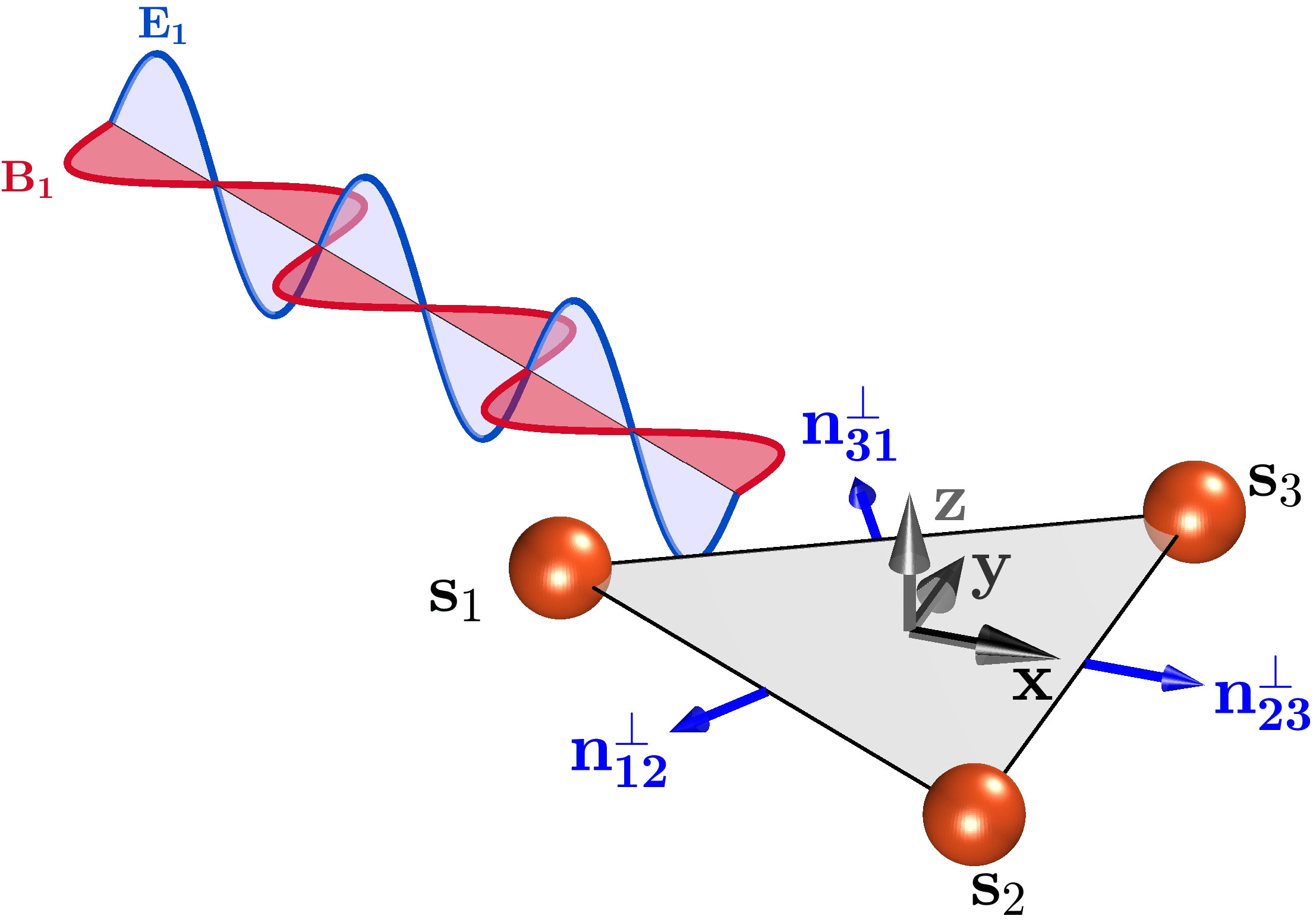}
\caption{Schematic view of a spin triangle interacting with a freely propagating beam. The scheme indicates the convention used in the calculations for the molecular reference frame (gray) and the interspin in-plane unit vectors ${\bf n}_{ij}^\perp $ (blue), perpendicular to the sides of the triangle.}
\label{fig0}
\end{figure}
Different mechanisms can in principle be responsible for a spin-electric coupling in molecular nanomagnets. At the spin-Hamiltonian level, the transitions between spin states induced by electric fields can result from the modulation --- mediated by spin-orbit interaction --- of different terms and parameters, ranging from the $g$ tensors or single-ion anisotropy, to the antisymmetric exchange \cite{Troiani19a,Yu22a}. A mechanism that is not mediated by spin-orbit coupling consists in the modulation of isotropic exchange couplings \cite{Trif08a,Trif10a}. This results from the fact that the relative alignment of two neighboring spins affects the charge distribution on the corresponding ions and on the interconnecting superexchange bridge. The relevance of this mechanism in molecular spin triangles based on transition metal ions was confirmed by ab initio calculations \cite{Islam10a,Nossa23a,Islam24a,Yu22a}, and by the successful interpretation of different experiments \cite{leMardele25a,Cini25a}.
In the following, we thus assume that the electric field essentially affects the spin cluster through the modulation of the isotropic exchange, and derive the electric-dipole operator accordingly.

In the observation of spin-electric coupling, a particular interest is devoted to antiferromagnetic spin triangles \cite{Choi06a,Ferrer12a,Boudalis21a,Boudalis18a,Robert19a,Liu19a,Kintzel21a,Lewkowitz23a,Cini25a,Singh2025Electrical}. There, in fact, spin frustration results in the formation of a ground $S=1/2$ quadruplet, whose states are defined by the total spin projection and by an additional degree of freedom (pseudospin). This is determined by the interaction that induces the zero-field splitting, which typically coincides with Dzyaloshinskii-Moriya, with inhomogeneous exchange couplings, or with a combination of the two. In all cases, an electric-field induced modulation of the exchange couplings can induce transitions between states with different values of the pseudospin, and identical values of the total-spin projection. A particularly promising case is that of the undistorted triangle with Dzyaloshinskii-Moriya interaction. Here, the pseudospin can be identified with the scalar chirality \cite{Bulaevskii08a,Trif08a,Trif10a,Georgeot10a,Belinsky14a}, and the spin-electric manipulation can be performed within subspaces that are intrinsically protected from hyperfine interactions \cite{Troiani12a}.

Here we extensively characterize both magnetic- and electric-dipole induced transitions in terms of the required field polarization. This is derived from the matrix elements of the magnetic and electric dipoles between eigenstates of prototypical spin triangles. In the presence of a well defined quantization axis, magnetic-dipole induced transitions require circularly-polarized light, as in the case of mononuclear paramagnetic systems \cite{Nehrkorn15a}. The polarization required for electric-dipole induced transitions, instead, varies from linear to circular, depending on whether the zero-field splitting in the ground multiplet originates from the triangle distorsion or on spin-orbit coupling. From this it follows that polarization can be used to discriminate not only between magnetic- and electric-dipole induced transitions, but also between different terms in the spin Hamiltonian. The investigation is extended to different spin lengths ($s_i=1/2, 3/2, 5/2$), and includes an analysis of relevant experimental geometries (Faraday and Voigt), defined by specific orientations of the field propagation direction relative to the spin-triangle and to the magnetic-field orientations.

The rest of the paper is organized as follows. In Section II we derive the polarizations that are required to induce the magnetic- and electric-dipole transitions in different kinds of spin triangles. Section III is devoted to the derivation of the polarization components that are relevant specifically in the Faraday and Voigt geometries. Our conclusions are reported in Section IV. Finally, in Appendix A we explain how the required polarizations are derived from the numerical calculation of the dipole matrix elements.

\section{Dipole operators and matrix elements}

If the $g$ tensor is isotropic and identical for all the spins within the cluster, the magnetic-dipole operator reads:
\begin{gather}
    \hat{\bf\mu}_b = g \mu_B  \hat{\bf S} \equiv \alpha_b \hat{\bf S}\,.
\end{gather}

The expression of the electric-dipole operator depends on the spin-electric coupling mechanism. We consider hereafter a coupling that results from the renormalization of the exchange couplings \cite{Trif08a,Trif10a,Islam10a,Nossa23a,Islam24a}. In this case, one can associate an electric dipole to each of the bonds connecting two adjacent spins ${\bf s}_i$ and ${\bf s}_j$. Such dipole depends on the relative orientation of the spins, and can thus be written as a vector operator that is proportional to ${\bf s}_i \cdot {\bf s}_j$ and perpendicular to the bond direction ${\bf r}_i-{\bf r}_j$, being ${\bf r}_i$ and ${\bf r}_j$ the positions of the spins. 

One can always introduce a reference frame so that the three spins lie on the $xy$ plane, at the positions ${\bf r}_1=d(-1,0)$, ${\bf r}_2=\frac{d}{2}(1,-\sqrt{3})$, and ${\bf r}_3=\frac{d}{2}(1,\sqrt{3})$, being $d$ the distance between each two spins. The unit vectors normal to the three bonds thus are: ${\bf n}_{12}^\perp = -\frac{1}{2} (1,\sqrt{3})$, 
${\bf n}_{23}^\perp = (1,0)$, and ${\bf n}_{31}^\perp = \frac{1}{2} (-1,\sqrt{3})$ (Fig. \ref{fig0}). As a result, the expression of the electric-dipole operator 
$\hat{\bf\mu}_e=\alpha_e\sum_{i<j} {\bf n}_{ij}^\perp\, \hat{\bf s}_i \cdot \hat{\bf s}_j$ reads:

\begin{gather}
    \hat{\bf\mu}_e = 
    \alpha_e{\bf x} \left[\hat{\bf s}_2 \cdot \hat{\bf s}_3-\frac{1}{2}(\hat{\bf s}_1 \cdot \hat{\bf s}_2 + \hat{\bf s}_3 \cdot \hat{\bf s}_1)\right] \nonumber\\ + \alpha_e\frac{\sqrt{3}}{2} {\bf y} (\hat{\bf s}_3 \cdot \hat{\bf s}_1 - \hat{\bf s}_1 \cdot \hat{\bf s}_2) \,,
\end{gather}
where $\alpha_e$ is the spin-electric coupling constant.

The amplitude of a dipole-induced transition $|i\rangle \rightarrow |f\rangle$ is proportional to the scalar product between the matrix element of the dipole operator $\hat{\bf\mu}$ and the unit vector ${\bf e}$ that defines the field polarization. Therefore, the polarization that maximizes the amplitude $A_{i\rightarrow f} \propto |{\bf e} \cdot \langle f | \hat{\bf\mu}| i \rangle|$ corresponds to the complex conjugate of the dipole matrix element, normalized to one:
\begin{gather}
    {\bf e} = \frac{\langle f | \hat{\bf\mu}| i \rangle^*}{|\langle f | \hat{\bf\mu}| i \rangle|} = \frac{\langle i | \hat{\bf\mu}| f \rangle}{|\langle f | \hat{\bf\mu}| i \rangle|}\,.
\end{gather}
In the following, we thus compute the matrix elements of the magnetic- and electric-dipole operators between eigenstates of different spin Hamiltonians, defining chirality, partial-spin sum, and hybrid (exchange) qubits.

\subsection{Chirality qubit}

\begin{figure}
\centering
\includegraphics[width=0.4\textwidth]{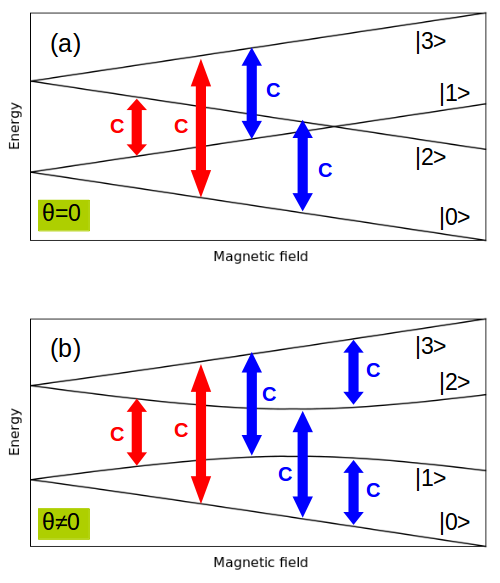}
\caption{Level scheme of the spin triangle with Dzyaloshinskii-Moriya interaction (chirality qubit), for magnetic field (a) oriented along the $z$ axis ($\theta=0$) or (b) tilted away from such axis ($\theta\neq 0$). Red and blue arrows correspond to magnetic- and electric-field induced transitions, respectively. The letter ``C'' denotes circular polarization.}
\label{figCQ1}
\end{figure}

The chirality qubit is defined by a spin triangle with a Hamiltonian $\hat{H}=\hat{H}_0+\hat{H}_c$, defined as follows \cite{Trif08a,Bulaevskii08a,Georgeot10a,Belinsky14a}:
\begin{gather}\label{ham_chirality}
    \hat{H}_0 = J \sum_{k=1}^3 \, \hat{\bf s}_k \cdot \hat{\bf s}_{k+1} + g\mu_B\,{\bf B}\cdot \hat{\bf S} \\
    \hat{H}_c = D\,{\bf z}\cdot \sum_{k=1}^3 \hat{\bf s}_k \times \hat{\bf s}_{k+1}  
    \,,
\end{gather}
with $ \hat{\bf s}_4 \equiv \hat{\bf s}_1 $. Here, $J>0$ and $D>0$ are the Heisenberg and Dzyaloshinskii-Moriya (DM) couplings, respectively. Typically, $J$ is much larger than $D$, so that the four lowest eigenstates belong to the $S=1/2$ quadruplet, on which we focus hereafter. In the case of $s_i=1/2$ spins, the four Hamiltonian eigenstates belonging to the $S=1/2$ subspace are also eigenstates of the scalar spin chirality
\begin{gather}
    \hat{\mathcal{C}}_z = \frac{4}{\sqrt{3}}\, \hat{\bf s}_1\cdot \hat{\bf s}_2 \times \hat{\bf s}_3 
\end{gather}
with eigenvalues $\pm 1$ (the normalization factor is included in order to have modulus one eigenvalues for $s_i=1/2$ spins, on which we mostly focus hereafter).

\subsubsection{Parallel magnetic field}

If the magnetic field is oriented along the positive $z$ direction, the four lowest Hamiltonian eigenstates $|\mathcal{C}_z,S_z\rangle$, belonging to the $S=1/2$ subspace and labeled after the values of the scalar chirality and of the total-spin projection, read \cite{Trif08a}:
\begin{gather}
|0\rangle \!\equiv\! |\!+\!1, \!-1/2 \rangle\! =\! \frac{1}{\sqrt{3}} ( |\!\uparrow\downarrow\downarrow\rangle \!+\! \epsilon |\!\downarrow\uparrow\downarrow\rangle \!+\! \epsilon^* |\!\downarrow\downarrow\uparrow\rangle )\label{eq06}\\
|1\rangle \!\equiv\! |\!-\!1,\!+1/2 \rangle\! =\! \frac{1}{\sqrt{3}} ( |\!\downarrow\uparrow\uparrow\rangle \!+\! \epsilon^* |\!\uparrow\downarrow\uparrow\rangle \!+\! \epsilon |\!\uparrow\uparrow\downarrow\rangle )\\
|2\rangle \!\equiv\! |\!-\!1,\!-1/2 \rangle\! =\! \frac{1}{\sqrt{3}} ( |\!\uparrow\downarrow\downarrow\rangle \!+\! \epsilon^* |\!\downarrow\uparrow\downarrow\rangle \!+\! \epsilon |\!\downarrow\downarrow\uparrow\rangle )\\
|3\rangle \!\equiv\! |\!+\!1, \!+1/2 \rangle \!=\! \frac{1}{\sqrt{3}} ( |\!\downarrow\uparrow\uparrow\rangle \!+\! \epsilon |\!\uparrow\downarrow\uparrow\rangle \!+\! \epsilon^* |\!\uparrow\uparrow\downarrow\rangle )\,,\label{eq07}
\end{gather}
where $\epsilon \equiv e^{i2\pi/3}=(-1+i\sqrt{3})/2$. 
The corresponding energy levels are [Fig. \ref{figCQ1}(a)]:
\begin{gather}
E_{0,3} = \mp \frac{1}{2}(\Delta_c+\Delta_b)\,,\ \ \ 
E_{1,2} = \mp \frac{1}{2}(\Delta_c-\Delta_b)\,,
\end{gather}
where $\Delta_b\equiv g\mu_B |{\bf B}|$ and $\Delta_c\equiv 2D$.
Hereafter, we assume for simplicity that $\Delta_c$ is larger than $\Delta_b$, so that $E_1$ is lower than $E_2$ and the labeling of the eigenstates reflects the ordering of the energy levels.  The chirality qubit is specifically defined by two states characterized by $S_z=-1/2$ and opposite values of $\mathcal{C}_z$. 

The nonzero matrix elements of the magnetic-dipole operator between the Hamiltonian eigenstates are:
\begin{gather}
\langle 0_c | \hat{\bf\mu}_b | 0_c \rangle = \langle 2_c | \hat{\bf\mu}_b | 2_c \rangle = -\frac{\alpha_b}{2}\,{\bf z}\label{eq10}\\
\langle 3 | \hat{\bf\mu}_b | 0 \rangle = \langle 2 | \hat{\bf\mu}_b | 1 \rangle^* = \frac{\alpha_b}{2}\,(i\,{\bf y}-{\bf x}) \equiv -\frac{\alpha_b}{\sqrt{2}}\,{\bf n}_-\\
\langle 3_c | \hat{\bf\mu}_b | 3_c \rangle = \langle 1_c | \hat{\bf\mu}_b | 1_c \rangle = +\frac{\alpha_b}{2}\,{\bf z}\,.\label{eq11}
\end{gather}
All other matrix elements are identically zero, as can be deduced from symmetry arguments. In fact, the total-spin components (and thus $\hat{\bf\mu}_b$) commute with the scalar chirality operator, and therefore can only have nonzero matrix elements between states with identical values of $\mathcal{C}_z$. From the above equations it follows that:
the transition $|0\rangle \rightarrow |3\rangle$ (characterized by $\Delta S_z=+1$ and by a constant scalar chirality $\mathcal{C}_z=+1$) is induced by radiation with left-handed circular polarization ${\bf e}_L = ({\bf x}+i{\bf y})/\sqrt{2}= {\bf n}_+$, being ${\bf e}_L$ parallel to $\langle 3|\hat{\bf\mu}_b|0\rangle^*$; 
the transition $|1\rangle \rightarrow |2\rangle$ (characterized by $\Delta S_z=-1$ and by a constant scalar chirality $\mathcal{C}_z=-1$) requires right-handed circular polarization ${\bf e}_R = ({\bf x}-i{\bf y})/\sqrt{2}= {\bf n}_-$, being ${\bf e}_R$ parallel to $\langle 2|\hat{\bf\mu}_b|1\rangle^*$.

Also the electric-field induced transitions require circular polarizations. 
In fact, the nonzero matrix elements of the electric dipole are:
\begin{gather}\label{eq:a}
\langle 2 | \hat{\bf\mu}_e | 0 \rangle = \langle 3 | \hat{\bf\mu}_e | 1 \rangle^* = \frac{3\alpha_e}{4}({\bf x}-i\,{\bf y}) \equiv \frac{3\alpha_e}{2\sqrt{2}}\,{\bf n}_-\,.
\end{gather}
All other matrix elements are identically zero. This follows from the fact that the exchange operators $\hat{\bf s}_i \cdot \hat{\bf s}_j$ (and thus $\hat{\bf\mu}_e$) commute with the total-spin projection operator, and therefore can only have have nonzero matrix elements between states with equal values of $S_z$. The diagonal matrix elements, however, are all zero, because 
$\langle\hat{\bf s}_1\cdot \hat{\bf s}_2\rangle = \langle\hat{\bf s}_2\cdot \hat{\bf s}_3\rangle = \langle\hat{\bf s}_3\cdot \hat{\bf s}_1\rangle = -1/4$, so that $\langle \hat{\mu}_e\rangle\propto\sum_{i<j} {\bf n}^\perp_{ij}=0$.
From Eq. (\ref{eq:a}) it follows that: 
the transition $|0\rangle \rightarrow |2\rangle$ ($\Delta C_z=-2$, total-spin projection $S_z=-1/2$) is induced by light with left-handed circular polarization ${\bf e}_L = {\bf n}_+$, being ${\bf e}_L$ parallel to $\langle 2|\hat{\bf\mu}_e|0\rangle^*$; 
the transition $|1\rangle \rightarrow |3\rangle$ ($\Delta C_z=+2$, total-spin projection $S_z=+1/2$) requires circular polarization ${\bf e}_R = {\bf n}_-$, being ${\bf e}_L $ parallel to $ \langle 3|\hat{\bf\mu}_e|1\rangle^*$. 


\subsubsection{Tilted magnetic field}

If the magnetic field ${\bf B}$ lies in the $xz$ plane and forms a nonzero angle $\theta$ with the $z$ axis, the scalar chirality still is a good quantum number. The Hamiltonian eigenstates are also characterized by well defined values of the spin projections $S_{\theta_\pm} \equiv S_z \cos\theta_\pm + S_x \sin\theta_\pm$, where  
\begin{gather}
    \theta_\pm \equiv \arctan \left(\frac{\Delta_b\sin\theta}{\Delta_b\cos\theta\pm \Delta_c}\right) \,,
\end{gather}
and the sign in the subscript corresponds to the (positive or negative) value of $\mathcal{C}_z$. The eigenstates can be expressed as follows in terms of those corresponding to the case $\theta=0$ [Eqs. (\ref{eq06}-\ref{eq07})]:
\begin{gather}
|0_\theta\rangle \!\equiv\! |\!+\!1, -1/2 \rangle \!=\! \cos (\theta_+/2) |0\rangle \!+\! \sin(\theta_+/2)|3\rangle \label{eq08}\\
|1_\theta\rangle \!\equiv\! |\!-\!1,+1/2 \rangle \!=\!  \cos (\theta_-/2) |1\rangle \!-\! \sin(\theta_-/2)|2\rangle \\
|2_\theta\rangle \!\equiv\! |\!-\!1,-1/2 \rangle \!=\!  \cos (\theta_-/2) |2\rangle \!+\! \sin(\theta_-/2)|1\rangle \\
|3_\theta\rangle \!\equiv\! |\!+\!1, +1/2 \rangle \!=\! \cos (\theta_+/2) |3\rangle \!-\! \sin(\theta_+/2)|0\rangle\,,\label{eq09}
\end{gather}
where the second quantum number that identifies the Hamiltonian eigenstate ($\pm 1/2$) refers to the value of $S_{\theta_+}$ and $S_{\theta_-}$ for positive ($|0_\theta\rangle$ and $|3_\theta\rangle$) and negative ($|1_\theta\rangle$ and $|2_\theta\rangle$) values of the scalar chirality, respectively. These two pairs of states thus form two Kramers doublets, but with two quantization axes that differ from each other, and from the axis that is parallel to the static magnetic field.

\begin{table}[]
\centering
\begin{tabular}{|c|c|c|c|}
\hline
Qubit & Transition  & Pol. & ${\bf e}$ \\
\hline
Chirality & $|0\rangle \xrightarrow{e} |2\rangle $  & C & $\frac{1}{\sqrt{2}}({\bf x}+i{\bf y})$ \\
($\theta=0$) & $|0\rangle \xrightarrow{m} |3\rangle $  & C & $\frac{1}{\sqrt{2}}({\bf x}+i{\bf y})$ \\
& $|1\rangle \xrightarrow{m} |2\rangle $  & C & $\frac{1}{\sqrt{2}}({\bf x}-i{\bf y})$ \\
& $|1\rangle \xrightarrow{e} |3\rangle $  & C & $\frac{1}{\sqrt{2}}({\bf x}-i{\bf y})$ \\
\hline
Chirality & $|0_\theta\rangle \xrightarrow{e} |1_\theta\rangle $  & C & $\frac{1}{\sqrt{2}}({\bf x}+i{\bf y})$ \\
($\theta\neq 0$) & $|0_\theta\rangle \xrightarrow{e} |2_\theta\rangle $  & C & $\frac{1}{\sqrt{2}}({\bf x}+i{\bf y})$ \\
& $|0_\theta\rangle \xrightarrow{m} |3_\theta\rangle $  & C & $\frac{1}{\sqrt{2}}({\bf x}\cos\theta_+-{\bf z}\sin\theta_++i{\bf y})$ \\
& $|1_\theta\rangle \xrightarrow{m} |2_\theta\rangle $  & C & $\frac{1}{\sqrt{2}}({\bf x}\cos\theta_--{\bf z}\sin\theta_--i{\bf y})$ \\
& $|1_\theta\rangle \xrightarrow{e} |3_\theta\rangle $  & C & $\frac{1}{\sqrt{2}}({\bf x}-i{\bf y})$ \\
& $|2_\theta\rangle \xrightarrow{e} |3_\theta\rangle $  & C & $\frac{1}{\sqrt{2}}({\bf x}-i{\bf y})$ \\
\hline
\end{tabular}
\caption{Summary of the polarizations required to address the electric- and magnetic-dipole induced transitions between the eigenstates ($S=1/2$ quadruplet) of the spin triangle ($s_i=1/2$) with Dzyaloshinskii-Moriya interaction (chirality qubit). The letters ``e'' and ``m'' denote the electric- and magnetic-dipole induced transitions, respectively. The last column indicates the polarization vector $\textbf{e}$.}
\label{tabCQ}
\end{table}

The corresponding energy levels are  [Fig. \ref{figCQ1}(b)]:
\begin{gather}
E_{0,3} = \mp \frac{1}{2}\left[(\Delta_c+\Delta_b\cos\theta)^2+(\Delta_b\sin\theta)^2\right]^{1/2}\\
E_{1,2} = \mp \frac{1}{2}\left[(\Delta_c-\Delta_b\cos\theta)^2+(\Delta_b\sin\theta)^2\right]^{1/2}\,.
\end{gather}
Hereafter we assume for simplicity that $0\le\theta\le\pi/2$ and $\Delta_b\le\Delta_c$, so that the labeling of the energy eigenvalues reflects their actual ordering.

The nonzero off-diagonal matrix elements of the magnetic-dipole operators can be derived from those obtained for $\theta=0$ [Eqs. (\ref{eq10}-\ref{eq11})], combined with the expressions of the eigenstates $|k_\theta\rangle$ [Eqs. (\ref{eq08}-\ref{eq09})]. Their expressions read:
\begin{gather}
\langle 3_\theta | \hat{\bf\mu}_b|0_\theta\rangle = \frac{\alpha_b}{2} ({\bf z}\sin\theta_+-{\bf x}\cos\theta_++i{\bf y})\label{eq01}\\
\langle 2_\theta | \hat{\bf\mu}_b|1_\theta\rangle = \frac{\alpha_b}{2} ({\bf z}\sin\theta_--{\bf x}\cos\theta_--i{\bf y})\,.\label{eq02}
\end{gather}
As for the case of a parallel magnetic field, the selection rule $\Delta\mathcal{C}_z=0$ applies, so that all the remaining off-diagonal elements are identically zero. From the above equations it follows that: the transition $|0_\theta\rangle \rightarrow |3_\theta\rangle$ (between states with scalar chirality $\mathcal{C}_z=+1$) is induced by light with left-handed circular polarization ${\bf e}_L = ({\bf z}\sin\theta_+-{\bf x}\cos\theta_+ - i{\bf y})/\sqrt{2}$, propagating along the direction ${\bf n}_+ = {\bf z}\,\cos\theta_+ + {\bf x}\,\sin\theta_+ $; the transition $|1_\theta\rangle \rightarrow |2_\theta\rangle$ (between states with scalar chirality $\mathcal{C}_z=-1$) requires right-handed circular polarization  ${\bf e}_R = ({\bf z}\sin\theta_--{\bf x}\cos\theta_- +i{\bf y})/\sqrt{2}$, propagating along the direction ${\bf n}_-={\bf z}\,\cos\theta_- + {\bf x}\,\sin\theta_- $. We note that neither ${\bf n}_+$ nor ${\bf n}_-$ are parallel to the magnetic field.

The nonzero off-diagonal matrix elements of the electric dipole are given by the expressions:
\begin{gather}
\langle 1_\theta | \hat{\bf\mu}_e|0_\theta\rangle \! =\! \langle 3_\theta | \hat{\bf\mu}_e|2_\theta\rangle^* \!=\! \frac{3\alpha_e}{4} ({\bf x}\!-\!i{\bf y}) \sin(\theta_+\!-\!\theta_-)\label{eq04}\\\!
\langle 2_\theta | \hat{\bf\mu}_e|0_\theta\rangle \! =\! \langle 3_\theta | \hat{\bf\mu}_e|1_\theta\rangle^* \!=\!  \frac{3\alpha_e}{4} ({\bf x}\!-\!i{\bf y}) \cos(\theta_+\!-\!\theta_-)\,.\label{eq05}
\end{gather}
From the above equations it follows that, independently on the orientation of the magnetic field: 
the circular polarization ${\bf e}_L$ is required for the transitions $|0_\theta\rangle \rightarrow |1_\theta\rangle$ and $|0_\theta\rangle \rightarrow |2_\theta\rangle$;
the circular polarization ${\bf e}_R$ is required for the transitions $|1_\theta\rangle \rightarrow |3_\theta\rangle$ and $|2_\theta\rangle \rightarrow |3_\theta\rangle$. 
However, the tilting angle $\theta$ and the zero-field splitting $\Delta_c$ determine the amplitudes of all these transitions, through the difference between the angles $\theta_+$ and $\theta_-$. In particular, a nonzero tilting angle allows the transitions $|0_\theta\rangle \rightarrow |1_\theta\rangle$ and $|2_\theta\rangle \rightarrow |3_\theta\rangle$, which are forbidden for $\theta=0$, and at the same time reduces the amplitudes of the transitions $|0_\theta\rangle \rightarrow |3_\theta\rangle$ and $|1_\theta\rangle \rightarrow |2_\theta\rangle$. This results from the fact that, for $\theta >0$, different quantization axes (associated with $S_{\theta_+}$ and $S_{\theta_-}$) are defined in the two chirality subspaces ($\mathcal{C}_z=+1$ and $\mathcal{C}_z=-1$). Therefore, the selection rules related to the conservation of the total spin projection only to transitions within a given subspace (and forbid the transitions $|0_\theta\rangle \rightarrow |3_\theta\rangle$ and $|1_\theta\rangle \rightarrow |2_\theta\rangle$), but not to inter-subspace transitions.

The polarizations required for the electric- and magnetic-induced transitions are summarized in Fig. \ref{figCQ1} and Table \ref{tabCQ}.

\subsection{Spin-sum qubit}

The spin-sum qubit is defined by a spin Hamiltonian $\hat{H}=\hat{H}_0+\hat{H}_{ss}$, where $\hat{H}_0$ is given in Eq. (\ref{ham_chirality}) and
\begin{gather}\label{ham_exchange}
    \hat{H}_{ss} = \Delta_J \, (\hat{\bf s}_1 \cdot \hat{\bf s}_2+1/4)\,.
\end{gather}
This term introduces an inhomogeneity between the exchange couplings. Without loss of generality, and simply to define an ordering of the eigenstates, we assume $\Delta_J>0$.
Typically, $J$ is much larger than $\Delta_J$, so that the four lowest eigenstates belong to the $S=1/2$ quadruplet, on which we focus hereafter.

For all values of the spin length $s_i$, the four lowest eigenstates of the above spin Hamiltonian are also eigenstates of the partial spin sum
$ \hat{\bf S}_{12}^2 = (\hat{\bf s}_1 + \hat{\bf s}_2)^2 $,
with eigenvalues $S_{12}(S_{12}+1)$ and $S_{12}=0,1$. The spin-sum qubit is specifically defined by two states characterized by the same value of the spin projection along the quantization axis and by different values of $S_{12}$.

\subsubsection{Parallel magnetic field}

\begin{figure}
\centering
\includegraphics[width=0.4\textwidth]{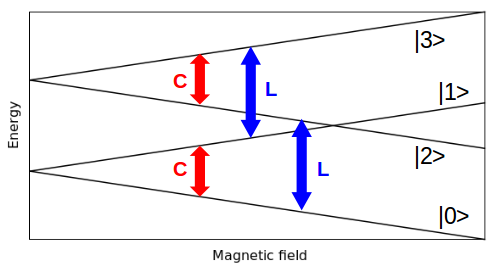}
\caption{Level scheme of the spin triangle with inhomogeneous Heisenberg couplings (spin-sum qubit), for arbitrary orientations of the magnetic field. Red and blue arrows correspond to magnetic- and electric-field induced transitions, respectively. The letter ``C'' (``L'') denotes circular (linear) polarization.}
\label{figPSSQ1}
\end{figure}

If $s_i=1/2$ and the magnetic field is oriented along the $z$ direction, the expressions of the four lowest eigenstates $|S_{12},S_z\rangle$, labelled by the values of the partial spin sum and of the total-spin projection, read:
\begin{gather}
|0\rangle \!\equiv\! |0, \!-1/2 \rangle\! =\! \frac{1}{\sqrt{2}} ( |\!\uparrow\downarrow\downarrow\rangle \!-\!  |\!\downarrow\uparrow\downarrow\rangle )\label{eq16}\\
|1\rangle \!\equiv\! |0, \!+1/2 \rangle\! =\! \frac{1}{\sqrt{2}} ( |\!\downarrow\uparrow\uparrow\rangle \!-\!  |\!\uparrow\downarrow\uparrow\rangle )\\
|2\rangle \!\equiv\! |1, \!-1/2 \rangle \!=\! \frac{1}{\sqrt{6}} ( |\!\uparrow\downarrow\downarrow\rangle \!+\! |\!\downarrow\uparrow\downarrow\rangle \!-\! 2
|\!\downarrow\downarrow\uparrow\rangle )\\
|3\rangle \!\equiv\! |1, \!+1/2 \rangle \!=\! \frac{1}{\sqrt{6}} ( |\!\downarrow\uparrow\uparrow\rangle \!+\! |\!\uparrow\downarrow\uparrow\rangle \!-\! 2
|\!\uparrow\uparrow\downarrow\rangle )\,.\label{eq17}
\end{gather}
The corresponding energy levels are:
\begin{gather}
E_{0,1} = -\frac{1}{2}(\Delta_J\pm\Delta_b)\,,\ \ \ 
E_{2,3} = \frac{1}{2}(\Delta_J\mp\Delta_b)\,,
\end{gather}
where $\Delta_B\equiv g\mu_B |{\bf B}|$ (Fig. \ref{figPSSQ1}). Hereafter, we assume for simplicity that $\Delta_J$ is larger than $\Delta_B$, so that $E_1$ is lower than $E_2$ and the labeling of the eigenstates reflects the ordering of the energy levels.

The nonzero matrix elements of the magnetic-dipole operator between the Hamiltonian eigenstates are:
\begin{gather}
\langle 0 | \hat{\bf\mu}_b | 0 \rangle = \langle 2 | \hat{\bf\mu}_b | 2 \rangle = -\frac{\alpha_b}{2}\,{\bf z}\label{eq12}\\
\langle 1 | \hat{\bf\mu}_b | 0 \rangle = \langle 3 | \hat{\bf\mu}_b | 2 \rangle = \frac{\alpha_b}{2}\,(i\,{\bf y}-{\bf x}) \equiv -\frac{\alpha_b}{\sqrt{2}}\,{\bf n}_-\,.\\
\langle 3 | \hat{\bf\mu}_b | 3 \rangle = \langle 1 | \hat{\bf\mu}_b | 1 \rangle = +\frac{\alpha_b}{2}\,{\bf z}\,.\label{eq13}
\end{gather}
All other matrix elements are identically zero, as can be deduced from symmetry arguments. In fact, the total-spin components (and thus $\hat{\bf\mu}_b$) commute with the exchange operators, and therefore can only have nonzero matrix elements between states with identical values of $S_{12}$. From the above equations it follows that: both the transitions $|0\rangle \rightarrow |1\rangle$ (between states with partial spin sum $S_{12}=0$) and $|2\rangle \rightarrow |3\rangle$ (between states with partial spin sum $S_{12}=1$) are induced by light with left-handed circular polarization ${\bf e}_L = ({\bf x}+i{\bf y})/\sqrt{2}$, being ${\bf e}_L$ parallel to $\langle 1 | \hat{\bf\mu}_b | 0 \rangle^* = \langle 3 | \hat{\bf\mu}_b | 2 \rangle^*$. 

The nonzero off-diagonal matrix elements of the electric dipole operator are:
\begin{gather}
\langle 2 | \hat{\bf\mu}_e | 0 \rangle = \langle 3 | \hat{\bf\mu}_e | 1 \rangle = \frac{3\alpha_e}{8}\,({\bf x}\sqrt{3} -{\bf y})\,,\label{eq20}
\end{gather}
and thus orthogonal to the diagonal matrix elements, {\it i.e.} to the expectation values 
\begin{gather}
\langle k | \hat{\bf\mu}_e | k \rangle = - \langle k' | \hat{\bf\mu}_e | k' \rangle = \frac{3\alpha_e}{8}\,({\bf x} +{\bf y}\,\sqrt{3})\,,\label{eq21}
\end{gather}
where $k=0,1$ and $k'=2,3$.
All other matrix elements are identically zero, because the exchange operators (and thus $\hat{\bf\mu}_e$) commute with the total spin components, and therefore can only have nonzero matrix elements between states with identical values of $S_z$. From Eq. (\ref{eq20}) it follows that the transitions $|0\rangle \rightarrow |2\rangle$ and $|1\rangle \rightarrow |3\rangle$ are induced by light with linear (``Horizontal'') polarization ${\bf e}_H = ({\bf x}\sqrt{3}-{\bf y})/2$, being ${\bf e}_H$ parallel to $\langle 2 | \hat{\bf\mu}_e | 0 \rangle^* = \langle 3 | \hat{\bf\mu}_e | 1 \rangle^*$.


\subsubsection{Tilted magnetic field}

The zero-field Hamiltonian of the spin-sum qubit is isotropic, so that the analysis performed for the case ${\bf B} \parallel {\bf z}$ does not change substantially for different orientations of the field in the $xz$ plane. However, we briefly report the relevant expressions for completeness.

The eigenstates can be labeled by the values of $S_{12}$ and of the spin projection along the field direction, $S_\theta\equiv S_z\cos\theta + S_x \sin\theta$. Their expressions read:
\begin{gather}
|0_\theta\rangle \!\equiv\! |0, \!-1/2 \rangle = \cos (\theta/2) \,|0\rangle + \sin(\theta/2)\,|1\rangle \label{eq14}\\
|1_\theta\rangle \!\equiv\! |0,\!+1/2 \rangle =  \cos (\theta/2) \,|1\rangle - \sin(\theta/2)\,|0\rangle \\
|2_\theta\rangle \!\equiv\! |1,\!-1/2 \rangle =  \cos (\theta/2) \,|2\rangle + \sin(\theta/2)\,|3\rangle \\
|3_\theta\rangle \!\equiv\! |1, \!+1/2 \rangle = \cos (\theta/2) \,|3\rangle - \sin(\theta/2)\,|2\rangle\,, \label{eq15}
\end{gather}
with the same energy levels reported for $\theta=0$ (Fig. \ref{figPSSQ1}).

The nonzero matrix elements of the magnetic dipole operator can be obtained by combining those obtained for $\theta=0$ [Eqs. (\ref{eq12}-\ref{eq13})] with the expressions of the eigenstates $|k_\theta\rangle$ [Eqs. (\ref{eq14}-\ref{eq15})]. The result is:
\begin{gather}
\langle 1_\theta | \hat{\bf\mu}_b|0_\theta\rangle \!=\! \langle 3_\theta | \hat{\bf\mu}_b|2_\theta\rangle \!=\! \frac{\alpha_b}{2} ({\bf z}\sin\theta\!-\!{\bf x}\cos\theta\!+\!i{\bf y})\,.\label{eq23}
\end{gather}
Both the transitions $|0\rangle \rightarrow |1\rangle$ and $|2\rangle \rightarrow |3\rangle$ are thus induced by light with left-handed circular polarization ${\bf e}_L = ({\bf x}\cos\theta\!-\!{\bf z}\sin\theta\!+\!i{\bf y})/\sqrt{2}$, propagating along the direction ${\bf n}={\bf z}\,\cos\theta + {\bf x}\,\sin\theta$. Transitions between states with different values of $S_{12}$ are forbidden, as for $\theta=0$.

The nonzero off-diagonal matrix elements of the electric dipole operator are:
\begin{gather}
\langle 2_\theta | \hat{\bf\mu}_e | 0_\theta \rangle = \langle 3_\theta | \hat{\bf\mu}_e | 1_\theta \rangle = \frac{3\alpha_e}{8}\,({\bf x}\sqrt{3} -{\bf y})\,.
\end{gather}
The transitions $|0_\theta\rangle \rightarrow |2_\theta\rangle$ and $|1_\theta\rangle \rightarrow |3_\theta\rangle$ thus require radiation with linear polarization ${\bf e}_H = ({\bf x}\sqrt{3}-{\bf y})/2$, propagating along any direction ${\bf n}$ perpendicular to ${\bf e}_H$, independently on the magnetic-field orientation. Transition between states with different values of $S_\theta$ are forbidden, given that $\hat{\bf\mu}_e$ and the total-spin projection operators commute.

The polarizations required for the electric- and magnetic-induced transitions are summarized in Fig. \ref{figPSSQ1} and Table \ref{tabPSSQ}.

\begin{table}[]
\centering
\begin{tabular}{|c|c|c|c|}
\hline
Qubit & Transition  & Pol. & ${\bf e}$ \\
\hline
Exchange & $|0\rangle \xrightarrow{m} |1\rangle $  & C & $\frac{1}{\sqrt{2}}({\bf x}+i{\bf y})$ \\
($\theta=0$) & $|0\rangle \xrightarrow{e} |2\rangle $  & L & $\frac{1}{2}({\bf x}\sqrt{3}-{\bf y})$ \\
& $|1\rangle \xrightarrow{e} |3\rangle $  & L & $\frac{1}{2}({\bf x}\sqrt{3}-{\bf y})$ \\
& $|2\rangle \xrightarrow{m} |3\rangle $  & C & $\frac{1}{\sqrt{2}}({\bf x}+i{\bf y})$ \\
\hline
Exchange & $|0_\theta\rangle \xrightarrow{m} |1_\theta\rangle $  & C & $\frac{1}{\sqrt{2}}({\bf x}\cos\theta-{\bf z}\sin\theta+i{\bf y})$ \\
($\theta\neq 0$) & $|0_\theta\rangle \xrightarrow{e} |2_\theta\rangle $  & L & $\frac{1}{2}({\bf x}\sqrt{3}-{\bf y})$ \\
& $|1_\theta\rangle \xrightarrow{e} |3_\theta\rangle $  & L & $\frac{1}{2}({\bf x}\sqrt{3}-{\bf y})$ \\
& $|2_\theta\rangle \xrightarrow{m} |3_\theta\rangle $  & C & $\frac{1}{\sqrt{2}}({\bf x}\cos\theta-{\bf z}\sin\theta+i{\bf y})$ \\
\hline
\end{tabular}
\caption{Summary of the polarizations required to address the electric- and magnetic-dipole induced transitions between the eigenstates ($S=1/2$ quadruplet) of the spin triangle ($s_i=1/2$) with inhomogeneous Heisenberg couplings (spin-sum qubit). The letters ``e'', ``m'', ``C'', and ``L'' denote electric- and magnetic-dipole induced transitions, circular and linear polarizations, respectively. }
\label{tabPSSQ}
\end{table}

\subsubsection{Scalene spin triangle}

The more general case where the three Heisenberg couplings are all different from one another can be dealt with, up to a term proportional to $\hat{\bf S}^2$ (which is irrelevant within the $S=1/2$ subspace), through the Hamiltonian term \cite{prova}:
\begin{gather}
    \hat{H}_{ss} \!=\! \Delta_J A_\varphi [\sqrt{3}\cos(2\varphi)\,\hat{\bf s}_1\!\cdot\! \hat{\bf s}_2\!+\sin(2\varphi)\,\hat{\bf s}_3\!\cdot\! (\hat{\bf s}_1\!-\!\hat{\bf s}_2)]\,.
\end{gather}
The prefactor $A_\varphi$ is defined as the inverse of the difference between the eigenstates of the term in square parentheses, and is such that the energy gap between the eigenstates of $\hat{H}_{ss}$ is always $\Delta_J$, independently on $\varphi$. The angle $\varphi$ defines the eigenstates through the expressions:
\begin{gather}
    |0_\varphi\rangle = \cos \varphi\, |0\rangle +\sin\varphi\, |2\rangle\label{eq18}\\
    |1_\varphi\rangle = \cos \varphi\, |1\rangle +\sin\varphi\, |3\rangle\\
    |2_\varphi\rangle = \cos \varphi\, |2\rangle -\sin\varphi\, |0\rangle\\
    |3_\varphi\rangle = \cos \varphi\, |3\rangle -\sin\varphi\, |1\rangle\,,\label{eq19}
\end{gather}
where the components $|k\rangle$ are defined in Eqs. (\ref{eq16}-\ref{eq17}). The above states $|0_\varphi\rangle$ and $|1_\varphi\rangle$ ($|2_\varphi\rangle$ and $|3_\varphi\rangle$) correspond to $S_{12}=0$ ($S_{12}=1$) for $\varphi=0$, to $S_{31}=0$ ($S_{31}=1$) for $\varphi=\pi/3$, and to $S_{23}=0$ ($S_{23}=1$) for $\varphi=2\pi/3$. For generic values of $\varphi$ none of the partial spin sums is a good quantum number. However, as discussed below, the resulting qubit, defined by a pair of states with an identical value of $S_z$ (for example, $|0_\varphi\rangle$ and $|2_\varphi\rangle$), displays the same properties as the partial spin sum qubit.

In fact, the matrix elements of the magnetic dipole operator are the same as in the case $\varphi=0$ [Eqs. (\ref{eq12}-\ref{eq13})]. The same circular polarization (${\bf e}_L$) is thus required to induce the transitions $|0_\varphi\rangle \rightarrow |1_\varphi\rangle$ and $|2_\varphi\rangle \rightarrow |3_\varphi\rangle$.

The matrix elements of the electric dipole operator, obtained by combining Eqs. (\ref{eq18}-\ref{eq19}) with Eqs. (\ref{eq20}-\ref{eq21}), depend on the mixing angle $\varphi$ through the expression:
\begin{gather}
    \langle 2_\varphi | \hat{\bf\mu}_e |0_\varphi\rangle = \langle 3_\varphi | \hat{\bf\mu}_e |1_\varphi\rangle \nonumber\\= \frac{3\alpha_e}{4} \left[{\bf x}\cos(2\varphi+\pi/6)-{\bf y}\sin(2\varphi+\pi/6)\right]\,.
\end{gather}
In all cases, the electric-dipole transitions are thus induced by a field with linear polarization ${\bf e}_H={\bf x}\cos(2\varphi+\pi/6)-{\bf y}\sin(2\varphi+\pi/6)$. This is perpendicular to all the expectation values $\langle k_\varphi | \hat{\bf\mu}_e | k_\varphi\rangle = \pm \left[{\bf x}\sin(2\varphi+\pi/6)+{\bf y}\cos(2\varphi+\pi/6)\right]$, where the positive and negative sign apply to $k=0,1$ and $k=2,3$, respectively. 

The generalization to the case of a tilted magnetic field ($\theta\neq 0$) is obtained by combining the results of the last two sections. 


\subsection{Generalized exchange qubit}

The generalized exchange qubit results from an interplay between the Dzyaloshinskii-Moriya interaction and an inhomogeneous exchange\cite{leMardele25a}. Its Hamiltonian is given by $\hat{H}=\hat{H}_0+\hat{H}_{ge}$, where:
\begin{gather}
   \hat{H}_{ge}=\hat{H}_{ss}\cos\phi + \hat{H}_c \sin\phi\,.
\end{gather}
The angle $\phi$ thus provides the degree of hybridization between the chirality and the exchange components, which are recovered as special cases for $\phi=\pi/2$ and $\phi=0$, respectively.

\subsubsection{Parallel magnetic field}

The eigenstates can be labelled by the values of $S_z$ and of an additional quantum number, corresponding to the lowest ($-1$) and heighest ($+1$) eigenvalues of $H_{ge}$:
\begin{gather}
|0_\phi\rangle \!=\! |\!-\!1,-1/2\rangle \!=\! \cos (\phi/2) |0\rangle \!+\! i \sin(\phi/2) |2\rangle \\
|1_\phi\rangle \!=\! |\!-\!1,+1/2\rangle \!=\! \cos (\phi/2) |1\rangle \!-\! i \sin(\phi/2) |3\rangle \\
|2_\phi\rangle \!=\! |\!+\!1,-1/2\rangle \!=\! i\sin (\phi/2) |0\rangle \!+\! \cos(\phi/2) |2\rangle \\
|3_\phi\rangle \!=\! |\!+\!1,+1/2\rangle \!=\! -i\sin (\phi/2) |1\rangle \!+\! \cos(\phi/2) |3\rangle \,.
\end{gather}
The first quantum number ($\pm 1$) that appears in the ket notation specifies the eigenvalue of $2\hat{H}_{ge}/\Delta$, being $\Delta\equiv (\Delta_J^2+\Delta_c^2)^{1/2}$. The states $|k\rangle$ that appear on the right-hand side of the above equations are the eigenstates of the spin-sum qubit for $\theta=0$ [Eqs. (\ref{eq16}-\ref{eq17})] \cite{riprova}. 

\begin{figure}
\centering
\includegraphics[width=0.4\textwidth]{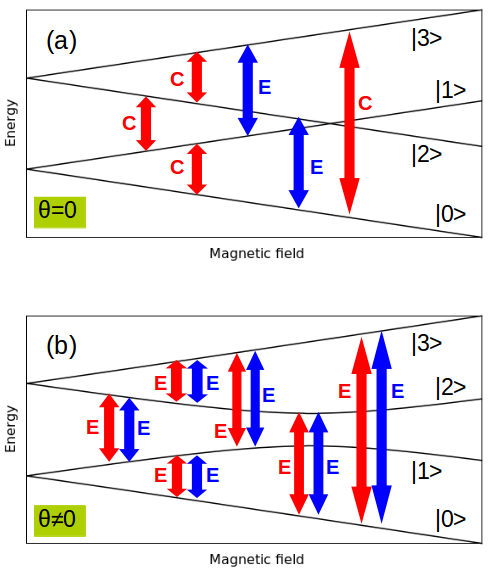}
\caption{Level scheme of the spin triangle with Dzyaloshinskii-Moriya interaction and inhomogeneous Heisenberg exchange (generalized exchange qubit), for magnetic field (a) oriented along the main molecule ($\theta=0$) or (b) tilted away from such axis ($\theta\neq 0$). Red (blue) arrows correspond to magnetic-field (electric-field) induced transitions. The letters ``C'', ``L'', and ``E'' denote circular, linear, and elliptical polarizations, respectively.}
\label{figHQ1}
\end{figure}

The energies corresponding to the states $|k_\phi\rangle$ are:
\begin{gather}
E_{0,1}=-\frac{1}{2}(\Delta\pm\Delta_b)\,,\ \ \ 
E_{2,3}=\frac{1}{2}(\Delta\mp\Delta_b)\,.
\end{gather}
The zero-field splitting $\Delta$ is related to those of the chirality and of the spin-sum qubits by the relations: $\Delta_c=\Delta\sin\phi$ and $\Delta_J=\Delta\cos\phi$.

The matrix elements of the magnetic dipole operator are given by the expressions:
\begin{gather}
    \langle 3_\phi | \hat{\bf\mu}_b | 2_\phi\rangle = \langle 1_\phi | \hat{\bf\mu}_b | 0_\phi\rangle = -\frac{\alpha_b}{\sqrt{2}} \,\cos\phi\,{\bf n}_-\\
    \langle 2_\phi | \hat{\bf\mu}_b | 1_\phi\rangle^* = \langle 3_\phi | \hat{\bf\mu}_b | 0_\phi\rangle = -\frac{i\alpha_b}{\sqrt{2}} \,\sin\phi\,{\bf n}_-\,.
\end{gather}
These transitions are thus induced by radiation with left-handed circular polarization, as for the chirality and for the spin-sum qubits. However, the selection rules related to $C_z$ and $S_{12}$ are removed for $0<\phi<\pi/2$, such that all four transitions between states with opposite values of $S_z$ are allowed.

\begin{table}[]
\centering
\begin{tabular}{|c|c|c|c|}
\hline
Qubit & Transition  & Pol. & ${\bf e}$ \\
\hline
Hybrid & $|0_\phi\rangle \xrightarrow{m} |1_\phi\rangle $  & C & $\frac{1}{2}({\bf x}+i{\bf y})$ \\
($\theta=0$) & $|0_\phi\rangle \xrightarrow{e} |2_\phi\rangle $  & E & $\frac{\xi}{2}[{\bf x}(\sqrt{3}+i\sin\phi)$ \\
& & & $-{\bf y}(1-i\sin\phi\sqrt{3})]$ \\
& $|0_\phi\rangle \xrightarrow{m} |3_\phi\rangle $  & C & $\frac{1}{2}({\bf x}+i{\bf y})$ \\
& $|1_\phi\rangle \xrightarrow{m} |2_\phi\rangle $  & C & $\frac{1}{2}({\bf x}-i{\bf y})$ \\
& $|1_\phi\rangle \xrightarrow{e} |3_\phi\rangle $  & E & $\frac{\xi}{2}[{\bf x}(\sqrt{3}-i\sin\phi)$ \\
& & & $-{\bf y}(1+i\sin\phi\sqrt{3})]$ \\
& $|2_\phi\rangle \xrightarrow{m} |3_\phi\rangle $  & C & $\frac{1}{2}({\bf x}+i{\bf y})$ \\
\hline
\end{tabular}
\caption{Summary of the polarizations required to address the electric- and magnetic-dipole induced transitions between the eigenstates ($S=1/2$ quadruplet) of the spin triangle ($s_i=1/2$) with Dzyaloshinskii-Moriya interaction and inhomogeneous Heisenberg couplings (generalized exchange qubit), for a magnetic field oriented along the $z$ axis. The letters ``e'' and ``m'' denote the electric- and magnetic-dipole induced transitions, respectively.}
\label{tabHQ1}
\end{table}

Electric dipole transitions are only allowed between states with identical value of $S_z$. The relevant matrix elements of the electric dipole operator read:
\begin{gather}
    \!\langle 2_\phi | \hat{\bf\mu}_e | 0_\phi\rangle \!=\! \frac{3\alpha_e}{8}\left[{\bf x} (\sqrt{3}\!-\!i\sin\phi)\!-\!{\bf y}(1\!+\!i\sin\phi\sqrt{3})\right]\\
    \!\!\!\!\langle 3_\phi | \hat{\bf\mu}_e | 1_\phi\rangle \!=\! \frac{3\alpha_e}{8}\left[{\bf x} (\sqrt{3}\!+\!i\sin\phi)\!-\!{\bf y}(1\!-\!i\sin\phi\sqrt{3})\right].
\end{gather}
These two transitions are induced by a field with elliptical polarization, defined by the unit vectors 
${\bf e}_+ \!=\! \frac{\xi}{2}\,[{\bf x} (\sqrt{3}\!+\!i\sin\phi)\!-\!{\bf y}(1\!-\!i\sin\phi\sqrt{3})]\!=\! \langle 2_\phi | \hat{\bf\mu}_e | 0_\phi\rangle^* / |\langle 2_\phi | \hat{\bf\mu}_e | 0_\phi\rangle|$
and
${\bf e}_- \!=\! \frac{\xi}{2}\,[{\bf x} (\sqrt{3}\!-\!i\sin\phi)\!-\!{\bf y}(1\!+\!i\sin\phi\sqrt{3})] \!=\! \langle 3_\phi | \hat{\bf\mu}_e | 1_\phi\rangle^* / |\langle 3_\phi | \hat{\bf\mu}_e | 1_\phi\rangle|$, where $\xi\equiv (1+\sin^2\phi)^{-1/2}$. From these equations one recovers the linear polarization if the exchange inhomogeneity is much larger than the DM coupling [$\phi=0$, Eq. (\ref{eq:a})], and the circular polarization in the opposite limit [$\phi=\pi/2$, Eq. (\ref{eq20})].

The polarizations required for the electric- and magnetic-induced transitions are summarized in Fig. \ref{figHQ1}(a) and Table \ref{tabHQ1}.

\subsubsection{Tilted magnetic field}

In the presence of both exchange inhomogeneity and DM interaction ($\phi\neq 0,\pi/2$) and for tilted magnetic fields ($\theta\neq 0$), simple analytical expressions of the dipole matrix elements cannot be provided. We thus characterize the transitions in terms of their intensities, normalized to the relevant coupling constant ($\chi=e,b$), 
\begin{gather}\label{eq:intensity}
 I \equiv |\langle k' | \hat{\bf\mu}_\chi | k\rangle |^2 / |\alpha_\chi|^2\,, 
\end{gather}
and of the degree of ellipticity $\eta\equiv \alpha/\beta$, defined as the ratio between the major and the minor axes of the ellipse (see Appendix A). The limiting cases $\eta=0$ and $\eta=1$ correspond to linear and circular polarizations, respectively. 

The values of $I$ and $\eta$ obtained for the generalized exchange qubit are reported in Table \ref{tabHQ2}, for different values of the mixing angle $\phi$. As a first remark, the interplay between the DM interaction and the inhomogeneous Heisenberg coupling, combined with the tilting of the magnetic field, removes all the selection rules that apply in the previous cases: neither the conservation of a scalar quantum number is required for the transitions induced by the magnetic dipole, nor the conservation of the spin projection is required for the transitions induced by the electric dipole. Both magnetic- and electric-dipole induced transitions are thus allowed ($I\neq 0$) for each pair of eigenstates. All transitions are characterized by an elliptical polarization ($0<\eta<1$), even if in some cases the polarization is nearly linear ($\eta$ is below $10^{-3}$, reported as 0$^*$ in the table). Concerning the distinguishability between the magnetic- and the electric-dipole transitions, we note the following. The absolute values of the intensities $I_\phi$ cannot be used directly for this purpose, because they do not include the value of the coupling constants, which are (at least in the electric case) generally unknown. However, one can rely, on the one hand, on the dependence of the intensities on the tilting angle $\theta$ and, on the other hand, on the polarization, which is always different for the electric- and magnetic-dipole induced transitions.

The polarizations required for the different transitions are summarized in Fig. \ref{figHQ1}(b).

\begin{table}[]
\centering
\begin{tabular}{|c|c|c|c|c|c|c|}
\hline
Transition  & $I_{\pi/8}$ & $\eta_{\pi/8}$ & $I_{\pi/4}$ & $\eta_{\pi/4}$ & $I_{3\pi/8}$ & $\eta_{3\pi/8}$ \\
\hline
$|0\rangle \xrightarrow{m} |1\rangle $  & $0.44$ & $0.96$ & $0.29$ & $0.86$ & $0.10$ & $0.76$  \\
$|0\rangle \xrightarrow{e} |1\rangle $ & $0^*$ & $0.28$ & $0.0019$ & $0.58$ & $0.0042$ & $0.86$ \\
$|0\rangle \xrightarrow{m} |2\rangle $  & $0.017$ & $0.10$ & $0.043$ & $0.10$ & $0.028$ & $0.10$ \\
$|0\rangle \xrightarrow{e} |2\rangle $  & $0.61$ & $0.28$ & $0.75$ & $0.58$ & $0.98$ & $0.86$ \\
$|0\rangle \xrightarrow{m} |3\rangle $  & $0.054$ & $0.76$ & $0.20$ & $0.84$ & $0.40$ & $0.94$ \\
$|0\rangle \xrightarrow{e} |3\rangle $  & $0.32$ & $0$ & $0.079$ & $0^*$ & $0.053$ & $0$ \\
$|1\rangle \xrightarrow{m} |2\rangle $  & $0.060$ & $0.70$ & $0.22$ & $0.79$ & $0.40$ & $0.92$ \\
$|1\rangle \xrightarrow{e} |2\rangle $  & $0.046$ & $0^*$ & $0.11$ & $0^*$ & $0.071$ & $0^*$ \\
$|1\rangle \xrightarrow{m} |3\rangle $  & $0.017$ & $0.10$ & $0.043$ & $0.10$ & $0.028$ & $0.10$ \\
$|1\rangle \xrightarrow{e} |3\rangle $  & $0.61$ & $0.28$ & $0.75$ & $0.58$ & $0.98$ & $0.86$ \\
$|2\rangle \xrightarrow{m} |3\rangle $  & $0.44$ & $0.96$ & $0.29$ & $0.86$ & $0.10$ & $0.76$ \\
$|2\rangle \xrightarrow{e} |3\rangle $  & $0^*$ & $0.28$ & $0.0019$ & $0.58$ & $0.0042$ & $0.86$ \\
\hline
\end{tabular}
\caption{Intensities ($I_\phi$) and ellipticities ($\eta_\phi$) of the electric- and magnetic-dipole induced transitions between the eigenstates ($S=1/2$ quadruplet) of the spin triangle ($s_i=1/2$) with Dzyaloshinskii-Moriya interaction and inhomogeneous Heisenberg couplings (generalized exchange qubit). The magnetic field is oriented along the $z$ axis tilted away from the $z$ axis ($\theta=\pi/4$). The symbol $0^*$ denotes numerical values that are between $10^{-4}$ and $10^{-3}$.}
\label{tabHQ2}
\end{table}

\subsection{Spins larger than 1/2}

We now consider triangles modeled through the same Hamiltonians $\hat{H}_c$, $\hat{H}_{ss}$, and $\hat{H}_{ge}$ considered above, but formed by larger spins, namely $s_i=3/2$ and $s_i=5/2$. Systems of this kind can be implemented through $\text{Cr}^{\text{III}}$- and $\text{Fe}^{\text{III}}$-based molecules, respectively \cite{robert_relevance_2019}. As for the triangles of $s_i=1/2$ spins, we focus on transitions between eigenstates belonging to the ground $S\approx 1/2$ quadruplet. 

One qualitative difference between the $s_i=1/2$ and $s_i>1/2$ cases is represented by the fact that, in the latter case, in the presence of DM interaction, neither the scalar chirality $\mathcal{C}_z$ nor the total spin $S$ are good quantum numbers, not even if the three exchange couplings are identical (see the Supplementary Information in Ref. \onlinecite{leMardele25a}). 

Let's start by considering the case  where the magnetic field is oriented along the main symmetry axis ($\theta = 0$). As reported in the Tables \ref{tabHQ3} and \ref{tabHQ4}, all the magnetic-dipole induced transitions are induced by a circular polarization, while the electric-dipole induced transitions require a polarization that evolves from linear ($\eta=0$) to circular ($\eta=1$), as the strength of the DM interaction ($\Delta_c$) increases relative to the inhomogeneity in the Heisenberg couplings ($\Delta_J$). 

When the magnetic field is tilted away from the $z$ axis ($\theta \neq 0$) and $\Delta_J,\Delta_c\neq 0$, all transitions within the ground quadruplet can be induced both by the electric and by the magnetic components of the field, by radiation with elliptical polarizations ($0<\eta<1$). If either $\Delta_J$ ($\phi=\pi/2$) or $\Delta_c$ ($\phi=0$) vanish, then electric-dipole induced transitions require linear or circular polarization, respectively, whereas magnetic-dipole induced transitions always require circular polarization  (Tables \ref{tabHQ3} and \ref{tabHQ4}). We finally note that, as the spin length $s_i$ increases, the intensities $I$ [Eq. (\ref{eq:intensity})]  of the magnetic transitions display a slight increase, due to the DM-induced mixing between $S=1/2$ and $S=3/2$ subspaces, while those related to electric transitions undergo a much more pronounced enhancement, for all orientations of the magnetic field. This latter aspect, however, cannot be directly related to experimental data, because the observed signal intensity, unlike $I$, also depends on the spin-electric coupling constant $\alpha_e$, which can vary significantly from one molecule to the other for any given value of the spin length $s_i$.

\section{Specific geometries}

Hereafter, we consider the components of the dipole operators that are relevant in specific experimental geometries, defined by the direction of the light propagation ${\bf n}$ with respect to the static magnetic field ${\bf B} = B(\sin\theta,0,\cos\theta) $. 
Given the propagation direction ${\bf n}$, the relevant part of the dipole moment for the transition $|i\rangle\rightarrow |f\rangle$ in the geometry $G$ reads:
\begin{gather}
\langle f| \hat{\bf\mu} |i\rangle_G=\langle f | \hat{\bf\mu} | i\rangle - ({\bf n} \cdot \langle f |\hat{\bf\mu}|i\rangle)\,{\bf n}\,.\label{eq22}
\end{gather}
In fact, the component of the dipole that is parallel to the field-propagation direction is perpendicular to the polarization vector ${\bf e}$, and thus provides no contribution to the amplitude $A_{i\rightarrow f}\propto|{\bf e} \cdot \langle f | \hat{\bf\mu}| i \rangle|$. The optimal polarization for this given field propagation direction is thus given by
\begin{gather}\label{eq:90}
    {\bf e} = \frac{\langle i | \hat{\bf\mu}| f \rangle_G}{|\langle f | \hat{\bf\mu}| i \rangle_G|}\,.
\end{gather}

From an opposite perspective, one might ask, given a dipole matrix element, which propagation direction maximizes the transition amplitude. The optimal ${\bf n}$ is the one that is orthogonal to the dipole matrix element. If $\mu_{fi}\equiv \langle f | \hat{\bf\mu}| i \rangle$ is a real vector, or can be reduced to a real vector by applying the same phase factor to the three spatial components ({\it i.e.} if the transition is induced by a linearly polarized field), then any of the infinite directions ${\bf n}$ that belong to the perpendicular plane is optimal. If instead the dipole matrix element is complex (circular or elliptical polarization), then one has a single optimal direction, namely the one that is perpendicular to both the real and imaginary components of $\mu_{fi}$.

\subsection{Faraday geometry}

In the Faraday geometry, the propagation direction is parallel to the magnetic field:
${\bf n} = {\bf n}_F \equiv (\sin\theta,0,\cos\theta)$. From this and from Eq. (\ref{eq22}) it follows that:
\begin{gather}
\langle f| \hat{\bf\mu} |i\rangle_F \!=\! {\bf\mu}_{fi}\! -\! (\mu_{fi,x}\sin\theta\!+\!\mu_{fi,z}z\cos\theta)\,{\bf n}_F\,,\label{eq03}
\end{gather}
being $\mu_{fi,x}$ and $\mu_{fi,z}$ the $x$ and $z$ components of ${\bf\mu}_{fi}$, respectively.

\subsubsection{Chirality qubit}

\begin{table}[]
\centering
\begin{tabular}{|c|c|c|c|c|c|c|}
\hline
Transition $(\theta=0)$ & $I_{0}$ & $\eta_{0}$ & $I_{\pi/4}$ & $\eta_{\pi/4}$ & $I_{\pi/2}$ & $\eta_{\pi/2}$ \\
\hline
$|0\rangle \xrightarrow{m} |1\rangle $ & $0.5$ & $1$ & $0.34$ & $1$ & $0$ & $-$ \\
$|0\rangle \xrightarrow{e} |2\rangle $ & $2.3$ & $0$ & $3$ & $0.57$ & $4.5$ & $1$ \\
$|0\rangle \xrightarrow{m} |3\rangle $ & $0$ & $-$ & $0.16$ & $1$ & $0.5$ & $1$ \\
$|1\rangle \xrightarrow{m} |2\rangle $ & $0$ & $-$ & $0.16$ & $1$ & $0.5$ & $1$ \\
$|1\rangle \xrightarrow{e} |3\rangle $ & $2.3$ & $0^*$ & $3$ & $0.57$ & $4.5$ & $1$ \\
$|2\rangle \xrightarrow{m} |3\rangle $ & $0.5$ & $1$ & $0.34$ & $1$ & $0$ & $-$ \\
\hline
Transition $(\theta=\pi/4)$ & $I_{0}$ & $\eta_{0}$ & $I_{\pi/4}$ & $\eta_{\pi/4}$ & $I_{\pi/2}$ & $\eta_{\pi/2}$ \\
\hline
$|0\rangle \xrightarrow{m} |1\rangle $ & $0.5$  & $1$       & $0.37$   & $0.91$   & $0$     & $-$    \\
$|0\rangle \xrightarrow{e} |1\rangle $ & $0$    & $-$       & $0.014$  & $0.44$   & $0.11$  & $1$    \\
$|0\rangle \xrightarrow{m} |2\rangle $ & $0$    & $-$       & $0.036$  & $0.17$   & $0$     & $-$    \\
$|0\rangle \xrightarrow{e} |2\rangle $ & $2.3$  & $0$       & $2.7$    & $0.44$   & $4.4$   & $1$    \\
$|0\rangle \xrightarrow{m} |3\rangle $ & $0$    & $-$       & $0.12$  & $0.82$   & $0.5$   & $1$ \\
$|0\rangle \xrightarrow{e} |3\rangle $ & $0$    & $-$       & $0.22$   & $0^*$ & $0$     & $-$    \\
$|1\rangle \xrightarrow{m} |2\rangle $ & $0$    & $-$       & $0.14$  & $0.71$   & $0.5$   & $1$ \\
$|1\rangle \xrightarrow{e} |2\rangle $ & $0$    & $-$       & $0.41$  & $0$      & $0$     & $-$    \\
$|1\rangle \xrightarrow{m} |3\rangle $ & $0$    & $-$       & $0.036$  & $0.17$   & $0$     & $-$    \\
$|1\rangle \xrightarrow{e} |3\rangle $ & $2.3$  & $0^*$       & $2.7$    & $0.44$   & $4.4$   & $1$    \\
$|2\rangle \xrightarrow{m} |3\rangle $ & $0.5$  & $1$       & $0.37$   & $0.91$   & $0$     & $-$    \\
$|2\rangle \xrightarrow{e} |3\rangle $ & $0$    & $-$       & $0.014$  & $0.44$   & $0.11$  & $1$    \\
\hline
\end{tabular}
\caption{Intensities ($I_\phi$) and ellipticities ($e_\phi$) of the electric- and magnetic-dipole induced transitions between the eigenstates (ground quadruplet) of the spin triangle ($s_i=3/2$) with Dzyaloshinskii-Moriya interaction and inhomogeneous Heisenberg couplings (generalized exchange qubit). The symbol $0^*$ denotes numerical values that are between $10^{-4}$ and $10^{-3}$.}
\label{tabHQ3}
\end{table}

\begin{table}[]
\centering
\begin{tabular}{|c|c|c|c|c|c|c|}
\hline
Transition $(\theta=0)$ & $I_{0}$ & $\eta_{0}$ & $I_{\pi/4}$ & $\eta_{\pi/4}$ & $I_{\pi/2}$ & $\eta_{\pi/2}$ \\
\hline
$|0\rangle \xrightarrow{m} |1\rangle $ & $0.5$ & $1$ & $0.25$ & $1$ & $0$ & $-$ \\
$|0\rangle \xrightarrow{e} |2\rangle $ & $5.1$ & $0$ & $7.8$ & $0.71$ & $10$ & $1$ \\
$|0\rangle \xrightarrow{m} |3\rangle $ & $0$ & $-$ & $0.26$ & $1$ & $0.52$ & $1$ \\
$|1\rangle \xrightarrow{m} |2\rangle $ & $0$ & $-$ & $0.26$ & $1$ & $0.52$ & $1$ \\
$|1\rangle \xrightarrow{e} |3\rangle $ & $5.1$ & $0$ & $7.8$ & $0.71$ & $10$ & $1$ \\
$|2\rangle \xrightarrow{m} |3\rangle $ & $0.5$ & $1$ & $0.25$ & $1$ & $0$ & $-$ \\
\hline
Transition $(\theta=\pi/4)$ & $I_{0}$ & $\eta_{0}$ & $I_{\pi/4}$ & $\eta_{\pi/4}$ & $I_{\pi/2}$ & $\eta_{\pi/2}$ \\
\hline
$|0\rangle \xrightarrow{m} |1\rangle $ & $0.5$ & $1$ & $0.29$ & $0.86$ & $0$ & $-$ \\
$|0\rangle \xrightarrow{e} |1\rangle $ & $0$ & $-$ & $0.018$ & $0.58$ & $0.052$ & $1$ \\
$|0\rangle \xrightarrow{m} |2\rangle $ & $0$ & $-$ & $0.044$ & $0.096$ & $0$ & $-$ \\
$|0\rangle \xrightarrow{e} |2\rangle $ & $5.1$ & $0^*$ & $6.9$ & $0.58$ & $10$ & $1$ \\
$|0\rangle \xrightarrow{m} |3\rangle $ & $0$ & $-$ & $0.21$ & $0.84$ & $0.52$ & $1$ \\
$|0\rangle \xrightarrow{e} |3\rangle $ & $0$ & $-$ & $0.74$ & $0.0033$ & $0$ & $-$ \\
$|1\rangle \xrightarrow{m} |2\rangle $ & $0$ & $-$ & $0.23$ & $0.79$ & $0.52$ & $1$ \\
$|1\rangle \xrightarrow{e} |2\rangle $ & $0$ & $-$ & $1.0$ & $0.0023$ & $0$ & $-$ \\
$|1\rangle \xrightarrow{m} |3\rangle $ & $0$ & $-$ & $0.044$ & $0.099$ & $0$ & $-$ \\
$|1\rangle \xrightarrow{e} |3\rangle $ & $5.1$ & $0^*$ & $6.9$ & $0.58$ & $10$ & $1$ \\
$|2\rangle \xrightarrow{m} |3\rangle $ & $0.5$ & $1$ & $0.29$ & $0.86$ & $0$ & $-$ \\
$|2\rangle \xrightarrow{e} |3\rangle $ & $0$ & $-$ & $0.017$ & $0.58$ & $0.052$ & $1$ \\
\hline
\end{tabular}
\caption{Intensities ($I_\phi$) and ellipticities ($e_\phi$) of the electric- and magnetic-dipole induced transitions between the eigenstates (ground quadruplet) of the spin triangle ($s_i=5/2$) with Dzyaloshinskii-Moriya interaction and inhomogeneous Heisenberg couplings (generalized exchange qubit). The symbol $0^*$ denotes numerical values that are between $10^{-4}$ and $10^{-3}$.}
\label{tabHQ4}
\end{table}

The component of the magnetic-dipole vector that is relevant in the Faraday geometry can be obtained by combining Eqs. (\ref{eq01}-\ref{eq02}) with the above Eq. (\ref{eq03}). One thus obtains for the chirality qubit the expressions:
\begin{gather}
\langle 3_\theta | \hat{\bf\mu}_{b} | 0_\theta\rangle_F = \frac{\alpha_b}{2} \left\{{\bf x} [-\cos\theta_+\!-\sin\theta\,\sin(\theta_+\!-\theta)] \right.\nonumber\\ \left. + i{\bf y}+ {\bf z} [\sin\theta_+\!-\cos\theta\sin(\theta_+\!-\theta)]\right\}\\
\langle 2_\theta | \hat{\bf\mu}_{b,F} | 1_\theta\rangle_F = \frac{\alpha_b}{2} \left\{{\bf x} [-\cos\theta_-\!-\sin\theta\,\sin(\theta_-\!-\theta)] \right.\nonumber\\ \left. - i{\bf y} + {\bf z} [\sin\theta_-\!-\cos\theta\sin(\theta_-\!-\theta)]\right\}\,.
\end{gather}
The transitions $|0_\theta\rangle \rightarrow |3_\theta\rangle$ and $|1_\theta\rangle \rightarrow |2_\theta\rangle$ can thus be induced by a field with elliptical polarization ${\bf e}$, given by the complex conjugates of the above transition amplitudes, normalized to 1. The ellipticity results from the fact that the imaginary ($y$) component of the relevant dipole matrix element coincides with that of ${\bf\mu}_{fi}$, while the real ($xz$) component is reduced by an amount that depends on the magnetic-field orientation.

The component of the electric-dipole vector that is relevant in the Faraday geometry can be obtained by combining Eqs. (\ref{eq04}-\ref{eq05}) with the above Eq. (\ref{eq03}). One thus obtains for the chirality qubit the expressions:
\begin{gather}
\langle 1_\theta | \hat{\bf\mu}_{e} | 0_\theta\rangle_F = \frac{3\alpha_e}{4} \,\sin(\theta_+\!-\theta_-)\nonumber\\ \left( {\bf x} \cos^2\theta - i{\bf y}- {\bf z} \sin\theta\cos\theta\right)\\
\langle 2_\theta | \hat{\bf\mu}_{e} | 0_\theta\rangle_F = \frac{3\alpha_e}{4} \,\cos(\theta_+\!-\theta_-) \nonumber\\ \left( {\bf x} \cos^2\theta - i{\bf y}- {\bf z} \sin\theta\cos\theta\right)\,.
\end{gather}
As for the overall electric dipole vectors, one has that
$\langle 3_\theta | \hat{\bf\mu}_{e} | 2_\theta\rangle_F = \langle 1_\theta | \hat{\bf\mu}_{e} | 0_\theta\rangle_F^*$
and
$\langle 3_\theta | \hat{\bf\mu}_{e} | 1_\theta\rangle_F = \langle 2_\theta | \hat{\bf\mu}_{e} | 0_\theta\rangle_F^*$. All the above transitions can thus be induced by a field with elliptical polarization ${\bf e}$, given by their with Eq. (\ref{eq:90}).

\subsubsection{Spin-sum qubit}

In the case of the spin-sum qubit, the polarization plane for the magnetic-dipole induced transitions is always perpendicular to the light propagation direction: $\langle f | \hat{\bf\mu}_{b} | i \rangle_F = \langle f | \hat{\bf\mu}_{b} | i \rangle$. The transitions are induced by a field with circular polarization, given by the normalized complex conjugates of the dipole vectors in Eq. (\ref{eq23}).

Being the electric-dipole vectors oriented on the $xy$ plane, their relevant components in the case of a Faraday geometry read:
\begin{gather}
\langle 2_\theta | \hat{\bf\mu}_{e} | 0_\theta\rangle_F = \frac{3\alpha_e}{8} \nonumber\\ \left( {\bf x} \sqrt{3}\cos^2\theta - {\bf y}- {\bf z}\sqrt{3} \sin\theta\cos\theta\right)\,,
\end{gather}
and $\langle 3_\theta | \hat{\bf\mu}_{e} | 1_\theta\rangle_F = \langle 2_\theta | \hat{\bf\mu}_{e} | 0_\theta\rangle_F$. These transitions are thus induced by a field with a linear polarization, given by the combination of the above equation with Eq. (\ref{eq:90}).

\subsection{Voigt geometry}

In the Voigt geometry, the propagation direction ${\bf n} = {\bf n}_V \equiv (\cos\theta\cos\chi,\sin\chi,-\sin\theta\cos\chi)$ is perpendicular to the magnetic field. For each direction of the magnetic field ($\theta$), one has an infinite set of propagation directions, corresponding to different values of $0\le\chi <2\pi$. From the expression of ${\bf n}_V$ and from Eq. (\ref{eq22}) it follows that:
\begin{gather}
\langle f |\hat{\bf\mu} |i\rangle_V= {\bf\mu}_{fi} \!-\! (\mu_{fi,x}\cos\theta\cos\chi\!+\!\mu_{fi,y}\sin\chi\nonumber\\-\!\mu_{fi,z}\sin\theta\cos\chi)\,{\bf n}_V\,,
\end{gather}
being $\mu_{fi,x}$, $\mu_{fi,y}$, and $\mu_{fi,z}$ the $x$, $y$, and $z$ components of the dipole matrix element, respectively.

\subsubsection{Chirality qubit}

The relevant part of the magnetic-dipole operator corresponding to the transitions $|0_\theta\rangle \rightarrow |3_\theta\rangle$ and $|1_\theta\rangle \rightarrow |2_\theta\rangle$ reads:
\begin{gather}
\langle 3_\theta | \hat{\bf\mu}_{b} | 0_\theta\rangle_V = \frac{\alpha_b}{2} \left[{\bf x}(A_+\cos\theta\cos\chi-\cos\theta_+)  \right.\nonumber\\ \left. + {\bf y}(A_+\sin\chi+i)+ {\bf z} (\sin\theta_+-A_+\sin\theta\cos\chi) \right]\\
\langle 2_\theta | \hat{\bf\mu}_{b} | 1_\theta\rangle_V = \frac{\alpha_b}{2} \left[{\bf x}(A_-\cos\theta\cos\chi-\cos\theta_-)  \right.\nonumber\\ \left. + {\bf y}(A_-\sin\chi-i)+ {\bf z} (\sin\theta_--A_-\sin\theta\cos\chi) \right]\,,
\end{gather}
where
$A_\pm\equiv \cos\chi\cos(\theta_\pm-\theta)\mp i\sin\chi$. These transitions are induced by a field with elliptical polarization ${\bf e}$, given by the combination of the above equations with Eq. (\ref{eq:90}).

The relevant part of the electric dipole corresponding to the allowed transitions reads: 
\begin{gather}
\langle 1_\theta | \hat{\bf\mu}_{e} | 0_\theta\rangle_V = \frac{3\alpha_e}{4} \,\sin(\theta_+-\theta_-)[ {\bf x} (1 +\nonumber\\ B\cos\theta\cos\chi) \!+\!{\bf y} (B\sin\chi\!-\!i) \!-\! {\bf z} B\sin\theta\cos\chi] \\
\langle 2_\theta | \hat{\bf\mu}_{e} | 0_\theta\rangle_V = \frac{3\alpha_e}{4} \,\cos(\theta_+-\theta_-) [ {\bf x} (1 +\nonumber\\ \!B\cos\theta\cos\chi) \!+\!{\bf y} (B\sin\chi\!-\!i) \!-\! {\bf z} B\sin\theta\cos\chi]\,,
\end{gather}
where
$B\!\equiv\! i\sin\chi\!-\!\cos\chi\cos\theta$.
In addition, one has that
$\langle 3_\theta | \hat{\bf\mu}_{e} | 2_\theta\rangle_V = \langle 1_\theta | \hat{\bf\mu}_{e} | 0_\theta\rangle_V^*$
and
$\langle 3_\theta | \hat{\bf\mu}_{e} | 1_\theta\rangle_V = \langle 2_\theta | \hat{\bf\mu}_{e,V} | 0_\theta\rangle_V^*$. All these transitions are thus induced by light with elliptical polarization ${\bf e}$, whose expression is obtained by combining the above equations with Eq. (\ref{eq:90}). 

\subsubsection{Spin-sum qubit}

The relevant part of the magnetic-dipole operator for the transition $|0_\theta\rangle \rightarrow |1_\theta\rangle$ is given by:
\begin{gather}
\langle 1_\theta | \hat{\bf\mu}_{b} | 0_\theta\rangle_V = \frac{\alpha_b}{\sqrt{2}} [ {\bf x} (C\cos\chi-1)\cos\theta \nonumber\\ +{\bf y} (C\sin\chi\!+\!i) \!+\! {\bf z} (1-C\cos\chi)\sin\theta] \,,
\end{gather}
where $C\equiv \cos\chi - i\sin\chi$. In addition, $\langle 3_\theta | \hat{\bf\mu}_{b} | 2_\theta\rangle_V = \langle 1_\theta | \hat{\bf\mu}_{b} | 0_\theta\rangle_V$. In both cases, the transition is induced by a field with elliptical polarization ${\bf e}=\langle 0_\theta | \hat{\bf\mu}_b|1_\theta\rangle_V / |\langle 0_\theta | \hat{\bf\mu}_b|1_\theta\rangle_V|$. 

The relevant component of the electric-dipole vector for the transition $|0_\theta\rangle \rightarrow |2_\theta\rangle$ is given by:
\begin{gather}
\langle 2_\theta | \hat{\bf\mu}_{e} | 0_\theta\rangle_V = \frac{3\alpha_e}{8} [{\bf x}\, (D\cos\theta\cos\chi +\sqrt{3})  \nonumber\\  + {\bf y}\,(D\sin\chi-1)- {\bf z}\, D\sin\theta\cos\chi ]\,.
\end{gather}
where 
$D\equiv\sin\chi - \sqrt{3}\cos\theta\cos\chi$.
In addition, $\langle 3_\theta | \hat{\bf\mu}_{e} | 1_\theta\rangle_V = \langle 2_\theta | \hat{\bf\mu}_{e} | 0_\theta\rangle_V$. In both cases, the transition is induced by a field with linear polarization ${\bf e}=\langle 0_\theta | \hat{\bf\mu}_b|2_\theta\rangle_V / |\langle 0_\theta | \hat{\bf\mu}_b|2_\theta\rangle_V|$.

\section{Conclusions}

We have investigated magnetic- and electric-dipole transitions in spin triangles, induced by freely propagating light beams. This exploration provides detailed tools for the interpretation of experiments carried out in the quasi-optical regime, where both the electric and the magnetic components of the oscillating field interact with the spin system. Based on the hypothesis that the electric-field induced transitions essentially results from the modulation of the symmetric exchange couplings, we have worked out the probabilities of the magnetically and electrically induced transitions for a series of spin Hamiltonians, spin lengths, polarizations and field propagation directions relative to the magnetic field (Faraday and Voigt geometries). Overall, these results show the crucial role that the field polarization can play in identifying the physical origin of the zero-field splitting in triangular molecules and in discriminating between electric- and magnetic-field induced transitions.

More specifically, a clear difference is found between the chirality and the partial spin sum qubits, where the zero-field splitting is induced by Dzyaloshinskii-Moriya and by an inhomogeneous exchange, respectively. In the former case, magnetic-dipole induced transitions occur between states belonging to different doublets, the electric-dipole transitions increase from two to four as the static magnetic field passes from parallel to tilted, and the required field polarization is always circular. In the latter case, the magnetic-dipole induced transitions occur between states belonging to the same doublet, while the electric-field induced transitions are always two, induced by a linearly polarized field, and independent of the magnetic field orientation and intensity. In the presence of both Dzyaloshinskii-Moriya and inhomogeneous exchange, and for a tilted magnetic field, all the selection rules resulting from the conservation of either $S_z$, $\mathcal{C}_z$, or $S_{12}$ break down, allowing both magnetic- and electric-dipole induced transitions between any two states belonging to the ground $S=1/2$ quadruplet. However, different kinds of transitions are characterized by different elliptical polarizations, displaying different dependencies on the orientation of the static magnetic field. The results obtained for triangles formed by spins with length $s_i>1/2$ qualitatively reproduce those obtained for $s_i=1/2$, even though the matrix elements of the electric dipole grow faster with $s_i$ than those of the magnetic dipole.

\appendix

\section{Elliptical polarization}

The derivation from the polarization vector ${\bf e}$ of the parameters $\alpha$ and $\beta$, and thus of the ellipticity $\eta$, has been carried out as follows. For a given transition $|i\rangle\rightarrow |f\rangle$, we compute the dipole vector ${\bf\mu}_{fi} = \langle f | \hat{\bf\mu} | i \rangle$ and define the relevant (optimal) polarization through the equation ${\bf e}={\bf\mu}^*_{fi}/|{\bf\mu}_{fi}|$. The real and imaginary parts of ${\bf e}$ define a plane, for which we identify two mutually orthogonal vectors ${\bf u}\equiv {\rm Re}({\bf e})/|{\rm Re}({\bf e})|$ and ${\bf v}=\{{\rm Im}({\bf e})-[{\rm Im}({\bf e})\cdot {\bf u}] {\bf u}\}/[1-({\bf e}\cdot {\bf u})^2]$, and the corresponding components $e_u\equiv {\bf u} \cdot {\bf e}$ and $e_v\equiv {\bf v} \cdot {\bf e}$. From these, we derive four parameters, whose expressions recall those of the Stokes parameters for the electric field \cite{Jackson}:
\begin{gather}
   s_0 \equiv |e_u|^2+|e_v|^2=\alpha^2+\beta^2\label{eq:app1}\\
   s_1 \equiv |e_u|^2-|e_v|^2=(\alpha^2-\beta^2)\cos(2\gamma)\\
   s_2 \equiv 2\,{\rm Re} (e_u^*e_v) = (\alpha^2-\beta^2)\sin(2\gamma)\\
   s_3 \equiv 2\,{\rm Im} (e_u^*e_v) = \alpha \beta \label{eq:app2}\,,
\end{gather}
where $\alpha$ and $\beta$ are the lengths of the semi-major and semi-minor axes, respectively, and $\gamma$ is the angle between the $x$ and the major axis. 

The validity of the above equations can be shown as follows.
Denoting with ${\bf x} = {\bf u} \cos\gamma +{\bf v}\sin\gamma$ and ${\bf y}=-{\bf u} \sin\gamma +{\bf v}\cos\gamma$, the unit vectors that are parallel to the major and minor axes of the ellipse, such that ${\bf e}=\alpha\, {\bf x} + i\beta\, {\bf y}$,
the polarization becomes ${\bf e} = e_u\, {\bf u} + e_v\, {\bf v}$, where:
\begin{gather}
e_u = \alpha\cos\gamma\!-\!i\beta\sin\gamma\,,\ \ \ 
e_v = \alpha\sin\gamma\!+\!i\beta\cos\gamma\,.
\end{gather}
Equations (\ref{eq:app1}-\ref{eq:app2}) follow from the above expression of $e_u$ and $e_v$ in terms of $\alpha$ and $\beta$. 

\FCM{If one applies, through a pulsed magnetic field creates a linear superposition between the states $|-1/2\rangle$ and $|+1/2\rangle$ of an $s=1/2$ spin, and lets the superposition evolve freely as a function of time (Larmor precession), one obtains ($\hbar=1$):
\begin{gather*}
|\psi\rangle = \frac{1}{\sqrt{2}} (|-1/2\rangle + e^{-i\Delta_bt}|+1/2\rangle)\,.
\end{gather*}
The expectation value of the dipole operator reads:
\begin{gather*}
\langle\psi| \hat{\mu}_b |\psi\rangle = \frac{1}{2} (\langle -1/2 |\hat{\mu}_b|+1/2\rangle + \langle -1/2 |\hat{\mu_e}|+1/2\rangle e^{-i\Delta_bt}+\nonumber\\ \langle +1/2 |\hat{\mu}_b|-1/2\rangle e^{i\Delta_bt}+\langle +1/2 |\hat{\mu}_b|+1/2\rangle) = \nonumber\\
\alpha_b[{\bf x}\cos(\Delta_bt)+{\bf y}\sin (\Delta_bt)]\,.
\end{gather*}
One thus obtains a rotating magnetic dipole.}

\FCM{If one applies, through a pulsed electric field creates a linear superposition between the states $|0\rangle$ (Eq. 7) and $|2\rangle$ (Eq. 9) of the chirality qubit, and lets the superposition evolve freely as a function of time (in analogy with the Larmor precession of a $1/2$ spin in a magnetic field), one obtains ($\hbar=1$):
\begin{gather*}
|\psi\rangle = \frac{1}{\sqrt{2}} (|0\rangle + e^{-i\Delta_ct}|2\rangle)\,.
\end{gather*}
The expectation value of the dipole operator reads:
\begin{gather*}
\langle\psi| \hat{\mu}_e |\psi\rangle = \frac{1}{2} (\langle 0 |\hat{\mu}_e|0\rangle + \langle 0 |\hat{\mu}_e|2\rangle e^{-i\Delta_ct}+\nonumber\\ \langle 2 |\hat{\mu}_e|0\rangle e^{i\Delta_ct}+\langle 2 |\hat{\mu}_e|2\rangle) = \nonumber\\
\frac{3\alpha_e}{4}[{\bf x}\cos(\Delta_ct)+{\bf y}\sin (\Delta_ct)]\,.
\end{gather*}
One thus obtains a rotating electric dipole.}

\acknowledgements 
We acknowledge financial support from the Italian Ministero dell’Universit\`a e della Ricerca (MUR), under  PNRR Project PE0000023-NQSTI, and from  the French Agence Nationale de la Recherche (ANR), under grant ANR-24-CE29-2752-01 (project SPINCHIRAL). Stimulating discussions with Milan Orlita and Jason McCoombs are also gratefully acknowledged. 


\begin{thebibliography}{42}
\bibitem [{rip()}]{riprova}%
  \BibitemOpen
  \href@noop {} {}\bibinfo {note} {The $|k_\phi\rangle$ coincide with the $|k\rangle$ for $\phi=0$, and (up to a phase factor) with the eigenstates of the chirality qubit for $\phi=\pi/2$. In particular, one has that: $|0_{\pi/2}\rangle = -i\epsilon |0_c\rangle$, $|2_{\pi/2}\rangle = -\epsilon^* |2_c\rangle$, $|1_{\pi/2}\rangle = i\epsilon^* |1_c\rangle$, and $|3_{\pi/2}\rangle = -\epsilon |3_c\rangle$.}\BibitemShut {Stop}%
\end{thebibliography}

%

\end{document}